\documentclass[9pt,journal]{IEEEtran}

\usepackage{kpfonts}
\usepackage{seqsplit}
\usepackage{placeins}
\usepackage{booktabs}
\usepackage[labelfont=bf]{caption}
\usepackage{nameref}
\usepackage{rotating}
\usepackage{float}
\usepackage{graphicx}
\usepackage{multirow}
\usepackage{newfloat}
\usepackage{hyperref}
\usepackage{vcell}
\usepackage{enumerate}
\hypersetup{colorlinks,citecolor=blue,linkcolor=blue,urlcolor=blue}
\usepackage{array}
\newcolumntype{R}[1]{>{\raggedright\arraybackslash}p{#1}}
\newcolumntype{C}[1]{>{\centering\let\newline\\\arraybackslash\hspace{0pt}}m{#1}}
\usepackage{setspace}
\usepackage{amsmath}
\usepackage{algorithm}
\usepackage{algorithmic}
\usepackage{subcaption}
\usepackage{multicol}
\usepackage{xcolor}
\usepackage{soul}
\usepackage{xspace}
\usepackage[most]{tcolorbox}
\definecolor{lightgreen}{rgb}{0.7, 0.9, 0.7} 
\sethlcolor{lightgreen}

\newcommand{\proposed}{GraphGDel\xspace}

\begin{document}

\title{GraphGDel: Constructing and Learning Graph Representations of Genome-Scale Metabolic Models for Growth-Coupled Gene Deletion Prediction}

\author{%
\IEEEauthorblockN{Ziwei Yang$^{1}$, Takeyuki Tamura$^{1}$}\\
\IEEEauthorblockA{\textit{$^{1}$ Bioinformatics Center, Institute for Chemical Research, Kyoto University}, Kyoto, Japan}%
\thanks{}}

\markboth{}%
{Shell \MakeLowercase{\textit{et al.}}: A Sample Article Using IEEEtran.cls for IEEE Journals}

\maketitle

\begin{abstract}
In genome-scale constraint-based metabolic models, gene deletion strategies are essential for achieving growth-coupled production, where cell growth and target metabolite synthesis occur simultaneously.
Despite the inherently networked nature of genome-scale metabolic models, existing computational approaches rely primarily on sequential data and lack graph representations that capture their complex relationships, as both well-defined graph constructions and learning frameworks capable of exploiting them remain largely unexplored.
To address this gap, we present a twofold solution. 
First, we introduce a systematic pipeline for constructing graph representations from constraint-based metabolic models. 
Second, we develop a deep learning framework that integrates these graph representations with gene and metabolite sequence data to predict growth-coupled gene deletion strategies. 
Across three metabolic models, our approach consistently outperforms established baselines, with improvements in overall accuracy of 14.04\%, 16.26\%, and 13.18\% over a deep feedforward neural network baseline, 6.17\%, 4.96\%, and 5.31\% over a sequence-learning baseline, and 5.10\%, 4.36\%, and 4.70\% over a topology-aware graph aggregation baseline on the same metabolite graph, respectively.
The source code and example datasets are available at: \url{https://github.com/MetNetComp/GraphGDel}.

\end{abstract}

\begin{IEEEkeywords}
Biology and genetics, Machine learning, Graphs and networks, Scientific databases.
\end{IEEEkeywords}

\section{INTRODUCTION}\label{sec:introduction}
Computational approaches are central to metabolic engineering, enabling the systematic design and optimization of microbial strains \cite{otero2021automated, garcia2021metabolic, foster2021building, toya2013flux, pharkya2006optimization}. 
A key application is computational strain design, which employs mathematical models to simulate cellular metabolism and guide the optimization of target metabolite biosynthesis.
Among available modeling frameworks, the \textbf{constraint-based model} is widely used in genome-scale metabolic engineering. 
It comprises two key components: (1) a metabolic network and (2) gene--protein--reaction (GPR) rules. 
The metabolic network defines the biochemical transformations among metabolites, catalyzed by enzymes encoded by specific genes. GPR rules, typically expressed as Boolean functions, describe the logical relationships between genes, proteins, and reactions, specifying whether a reaction is active based on gene presence. 
By altering gene content, the metabolic network can be systematically reprogrammed. 
One widely applied operation is \textbf{gene deletion}, which disables specific reactions by removing genes that encode essential enzymes. 
This approach, central to metabolic engineering, enables the rational rewiring of metabolism to enhance the biosynthesis of target compounds.

In constraint-based simulations of gene deletions, the objective is not merely to increase the production of a target metabolite, but to achieve \textbf{growth-coupled production}---a state in which the synthesis of the target metabolite is coupled with cellular growth.
This coupling is critical for industrial biotech applications, as microorganism strains with higher growth rates are more likely to dominate during repeated passaging, thereby ensuring stable and sustained metabolite production over time.
However, under native metabolic conditions, only a limited number of metabolites naturally exhibit growth-coupling. 
To extend this property to a broader range of target compounds, it is necessary to design \textbf{gene deletion strategies} that reconfigure the metabolic network to enforce growth-coupled production~\cite{vieira2019comparison}. 

To this end, a range of computational methods have been developed over the past decade to identify gene deletion strategies capable of enabling growth-coupled phenotypes~\cite{yang2011emilio,egen2012truncated,rockwell2013redirector,ohno2014fastpros,gu2016idealknock,tamura2018grid,tamura2023gene}.
These strategies have been successfully applied in real-world contexts, facilitating the development of engineered strains for applications, including biofuel production and the industrial biosynthesis of platform chemicals~\cite{trinh2008minimal,schneider2020extended,banerjee2020genome}.
Despite these advances, calculating gene deletion strategies \textit{de novo} remains computationally intensive, particularly for genome-scale metabolic models that encompass complex network structures and GPR associations involving numerous genes.
In response, recent efforts have focused not only on developing more mathematically efficient algorithms to accelerate these computations, but also on establishing databases to store and manage strain design data, thereby enabling more scalable and effective solutions.
One prominent example is \textit{MetNetComp} database~\cite{tamura2023metnetcomp}, the first web-based platform to curate gene deletion strategies for growth-coupled production within constraint-based metabolic models. 
\textit{MetNetComp} extends the \textit{gDel\_minRN} method~\cite{tamura2023gene} to compute gene deletion strategies that enable growth coupling, supporting both minimal and maximal gene deletion configurations. 
Currently, it hosts a comprehensive repository of over 85{,}000 gene deletion strategies spanning a wide range of metabolites and multiple constraint-based models from diverse species. 
Such resources are essential not only for meeting the growing demand for gene deletion analysis but also for enabling new paradigms in strain design, including data-driven approaches to gene deletion strategy development.

Designing data-driven approaches for gene deletion strategy development remains a non-trivial task, with a central challenge being the effective extraction and utilization of information from existing gene deletion strategy databases.
The first effort in this direction, \textit{DBgDel}~\cite{yang2025dbgdel}, addressed this challenge by summarizing known maximal gene deletion strategies from the database and using them as constraints to limit the search space during optimization.
This method strikes a favorable balance between computational efficiency and success rate when applied to genome-scale metabolic models.
However, \textit{DBgDel} relies on hand-crafted extraction rules, limiting its ability to capture complex patterns from large-scale datasets, and still depends on downstream optimization algorithms to properly incorporate GPR rules.
To overcome these limitations, \textit{DeepGDel}~\cite{yang2025deepgdel} was introduced as the first framework to formulate gene deletion strategy design as a predictive learning task using deep learning techniques~\cite{lecun2015deep}, laying the foundation for fully data-driven gene deletion strategy design.
\textit{DeepGDel} jointly models gene and metabolite sequences along with existing gene deletion strategy data, enabling end-to-end prediction of growth-coupled gene deletion strategies and demonstrating robust applicability across metabolic models from diverse species and scales.
Although \textit{DeepGDel} effectively models metabolite features from sequential input data, it ignores upstream and downstream interactions that are widely recognized as essential for understanding the complexity of metabolic systems \cite{structure_1, structure_2, structure_3}. 
This limitation reduces predictive performance, particularly on large genome-scale models. 
Graph-based representations provide a natural framework for capturing such structured dependencies and offer critical context for understanding the coupling between target metabolite production and cell growth. 
Recent advances in graph learning have demonstrated strong potential for extracting meaningful relationships from graph-structured biological data, enabling effective solutions across diverse biological tasks \cite{graphlearning_1,graphlearning_2,graphlearning_3}.
Based on these observations, integrating graph-based learning into gene deletion strategy design to better capture metabolite interactions is highly desirable.

However, achieving this goal is not straightforward.
Unlike curated pathway maps, such as those provided by KEGG (Kyoto Encyclopedia of Genes and Genomes) \cite{KEGG}, genome-scale metabolic models do not explicitly define topological relationships among biological entities in the form of a graph \cite{orth2010flux,simeonidis2015genome}.
Instead, they rely on algebraic representations: stoichiometric matrices encoding mass-balance constraints, Boolean expressions defining GPR rules, and flux bounds specified as interval constraints.
Since these formulations are inherently algebraic rather than topological, they lack an explicit graph structure in which nodes represent biological entities and edges represent their interactions.
This absence of graph structure hinders the direct application of metabolic models in graph-based computation and learning.
Moreover, there is a central challenge in constructing meaningful graph representations from metabolic models: the treatment of ubiquitous metabolites, such as ATP, NAD(H), and other common cofactors that participate in a wide range of reactions.
These metabolite act as topological hubs, often obscuring the modular organization of metabolic networks and diluting biologically meaningful structures \cite{hubs,structure_3}.
Although such entities are commonly collapsed or excluded in databases like KEGG and BiGG to improve visual clarity, their impact on graph-based learning remains underexplored and poorly understood \cite{king2016bigg}.

To address the aforementioned challenges, this work aims to achieve two primary objectives. 
First, we propose a systematic approach to \textbf{construct graph representations} from input constraint-based metabolic models.
This construction process incorporates two core refinement steps.
The graph attribute-based refinement focuses on filtering and removing highly connected nodes, while the knowledge-based refinement targets the editing of currency metabolite nodes, thereby producing graphs that are both biologically grounded and informative.
Second, we introduce \textit{\proposed}, a deep learning framework that learning from both sequence data and constructed metabolic graphs. 
\textit{\proposed} consists of four neural network-based modules:  
(1) \textbf{Gene-M}, a sequence learning module that encodes gene inputs;  
(2) \textbf{Meta-M}, a sequence learning module that encodes metabolite inputs;  
(3) \textbf{Graph-M}, a graph learning module that further captures metabolite interactions within the metabolic model; 
and (4) \textbf{Pred-M}, a prediction module that integrates upstream representations to predict growth-coupled gene deletion strategies in genome-scale metabolic models.

To the best of our knowledge, this is the first study to explore graph learning in the context of gene deletion strategy prediction.
Computational experiments demonstrated that the graphs constructed by the proposed method are topologically simplified yet remain well-suited for graph-based learning tasks on genome-scale metabolic models.
Furthermore, the proposed framework, \textit{\proposed}, effectively predicted gene deletion strategies for growth-coupled production across three metabolic models in a fully data-driven manner, achieving robust performance and outperforming baseline methods in overall accuracy, macro-averaged precision, recall, F1 score, and AUC.

\section{PRELIMINARY AND PROBLEM SETTING}
\label{sec:Preliminary and Problem Setting}
In this section, we first outline several key terms in this study as a preliminary.
Next, we illustrate these concepts using a toy constraint-based model.
Finally, we define the main problem in this research, namely, \textbf{gene deletion strategy prediction}. 
All the notations mentioned in this section are listed in a table, which can be found in the Appendix.

\subsection{Preliminary Definition}
\label{subsec:Preliminary}

\subsubsection{\textbf{Constraint-based model}}
Let $\mathcal{C} = \{\textit{M}, \textit{R}, \textit{S}, \textit{L}, \textit{U}, \textit{G}, \textit{F}, \textit{P}\}$ be a constraint-based model.
The elements of $\mathcal{C}$ can be further divided into two parts: (1) the metabolic network and (2) the GPR rule. 
We describe them as follows:

\begin{itemize}
  \item 
  \textbf{Metabolic network}: Let $\mathcal{C}_1 = \{\textit{M}, \textit{R}, \textit{S}, \textit{L}, \textit{U}\}$ be a metabolic network.
  $\textit{M} = \{m_1, \ldots, m_{\left\lvert \textit{M} \right\rvert}\}$ denotes a set consisting all metabolites, with one of them being the target metabolite $m_{\text{target}}$.
  $\textit{R} = \{r_1, \ldots, r_{\left\lvert \textit{R} \right\rvert}\}$ denotes a set consisting all reactions, including the cell growth reaction $r_{\text{growth}}$ and the target metabolite production reaction $r_{\text{target}}$.
  To facilitate clarity, we introduce an additional set $\textit{V} = \{v_1, \ldots, v_{\left\lvert \textit{R} \right\rvert}\}$ to represent the reaction rates per unit time (flux) corresponding to reactions in $R$. 
  Notably, we distinguish the rate of the growth reaction ($v_{\text{growth}}$) as the \textbf{Growth Rate (GR)} and that of the target metabolite production reaction ($v_{\text{target}}$) as the \textbf{Production Rate (PR)}.
  The stoichiometry matrix $S$ contains elements $S_{ij} = k$, indicating that reaction $r_j$ either produces (+) or consumes (-) $k$ units of metabolite $m_i$ per unit time.
  Lower and upper bounds for the reaction rates for $\textit{V}$ are denoted by $\textit{L} = \{l_1, \ldots, l_{\left\lvert \textit{R} \right\rvert}\}$ and $\textit{U} = \{u_1, \ldots, u_{\left\lvert \textit{R} \right\rvert}\}$, respectively.

  \item
  \textbf{GPR rule}: Let $\mathcal{C}_2 = \{\textit{G}, \textit{F}, \textit{P}\}$ be a GPR rule.
  $\textit{G} = \{g_1, \ldots, g_{\left\lvert \textit{G} \right\rvert}\}$ represents a set of genes, while $\textit{F} = \{f_1, \ldots, f_{\left\lvert \textit{R} \right\rvert}\}$ represents Boolean functions.
  $\textit{P} = \{p_1, \ldots, p_{\left\lvert \textit{R} \right\rvert}\}$ is a collection of outputs generated by applying the functions in $\textit{F}$ to the gene set $\textit{G}$. 
  Each output $p_j$ can be expressed as $p_j = f_j(\textit{G})$, where both $p$ and $g$ are binary values, i.e., $p, g \in \{0,1\}$.
  Note that if $p_j = 0$, this imposes a constraint on both the lower bound $l_j$ and the upper bound $u_j$ to be 0, effectively restraining reaction $r_j$.
\end{itemize}

\subsubsection{\textbf{Flux balance analysis}}
When analyzing a metabolic network within a constraint-based model, flux balance analysis (FBA) assumes a steady state in which all metabolic reaction fluxes remain constant~\cite{orth2010flux}. 
Formally, FBA imposes the following conditions:  
(1) for each metabolite, the sum of producing fluxes equals the sum of consuming fluxes;  
(2) in each reaction, substrate and product fluxes must satisfy the stoichiometric ratios specified by the reaction equation; and  
(3) each flux is constrained by prescribed upper and lower bounds.  
In the standard formulation of FBA applied to a metabolic network $\mathcal{C}_1$, the objective is to maximize the cell growth reaction rate $v_{\text{growth}}$, expressed as the following linear program (LP):

\begin{algorithm}
  \caption*{LP Formalization of FBA Using $\mathcal{C}_1$ of a Constraint-Based Model.}
  \textbf{Given:} $\mathcal{C}_1$ \\
  \textbf{Maximize:} $v_{\text{growth}}$ \par\vspace{-0.5em}
  \begin{flushleft}
    \hspace{1em}\textbf{Such that:} \\
    \hspace{2em}$\sum_{j} S_{ij} v_{j} = 0 \quad \text{for all } i$;\\
    \hspace{2em}$l_j \leq v_j \leq u_j \quad \text{for all } j$;\\
    \hspace{2em}$i = \{1, \ldots, |\mathit{M}|\}, \quad j = \{1, \ldots, |\mathit{R}|\}$
  \end{flushleft}
  \label{algo:LP}
\end{algorithm}

\subsubsection{\textbf{Growth-coupled production}}
Growth-coupled production is a special case of microbial metabolism, wherein cell growth co-occurs with the synthesis of a target metabolite. 
In this study, we adopt the paradigm of \textit{weakly growth-coupled production}, which requires achieving a non-zero synthesis rate of the target metabolite at the maximum non-zero growth rate~\cite{schneider2021systematizing, alter2019determination}. 
During the simulation of constraint-based models, growth-coupled production can be characterized by two indices: 
the lower bound of the target metabolite production rate ($PR_{\text{threshold}}$) and the lower bound of the cell growth reaction rate ($GR_{\text{threshold}}$). 
When the growth rate is maximized, if the simulated values $v_{\text{growth}}$ and minimum $v_{\text{target}}$ exceed their respective lower bounds, growth-coupled production is considered to be achieved. 
In this study, we set $PR_{\text{threshold}} = GR_{\text{threshold}} = 0.001$.

\subsubsection{\textbf{Gene deletion strategy}}
In metabolic model simulations, a gene deletion strategy refers to selecting genes for removal in order to optimize the network toward desired outcomes, such as the growth-coupled production of specific metabolites. 
For growth-coupled production, the gene deletion strategy is typically formulated as identifying a set \( \mathit{G}_{\text{del}} \) of genes designated for deletion, as follows:

\begin{algorithm}
  \caption*{Calculation of Gene Deletion Strategy for Growth-Coupled Production.}
  \textbf{Given:} $\mathcal{C}$, $r_{\text{target}}$, $PR_{\text{threshold}}$, $GR_{\text{threshold}}$ \\
  \textbf{Find:} $\textit{G}_\text{del} \subset \textit{G}$ such that $v_{\text{target}} \geq PR_{\text{threshold}}$ and $v_{\text{growth}} \geq GR_{\text{threshold}}$ \par\vspace{-0.5em}
  \begin{flushleft}
    \hspace{1em}\textbf{Such that Minimize:} \\
    \hspace{2em}$v_{\text{target}}$ \\
    \hspace{2em}\textbf{Such that Maximize:} \\
    \hspace{3em}$v_{\text{growth}}$ \\
    \hspace{3em}\textbf{Such that:} \\
    \hspace{4em}$\sum_{j} S_{ij} v_{j} = 0 \quad \text{for all } i$; \\
    \hspace{4em}
    $\begin{cases}
      v_j = 0 & \text{if } p_j = 0, \\
      l_j \leq v_j \leq u_j & \text{otherwise};
    \end{cases}$ \\
    \hspace{4em}$p_j = f_j(\textit{G})$; \\
    \hspace{4em}
    $\begin{cases}
      g = 0 & \text{if } g \in \textit{G}_\text{del}, \\
      g = 1 & \text{otherwise};
    \end{cases}$
  \end{flushleft}
  \label{algo: Calculation of Gene Deletion Strategy}
\end{algorithm}

\subsection{Example: Gene Deletions in Metabolic Models}
\label{subsec:Example}
A small example with accompanying explanations is provided below to further illustrate gene deletions in metabolic models.

\begin{figure}[t]
  \centering
  \includegraphics[width=0.9\linewidth]{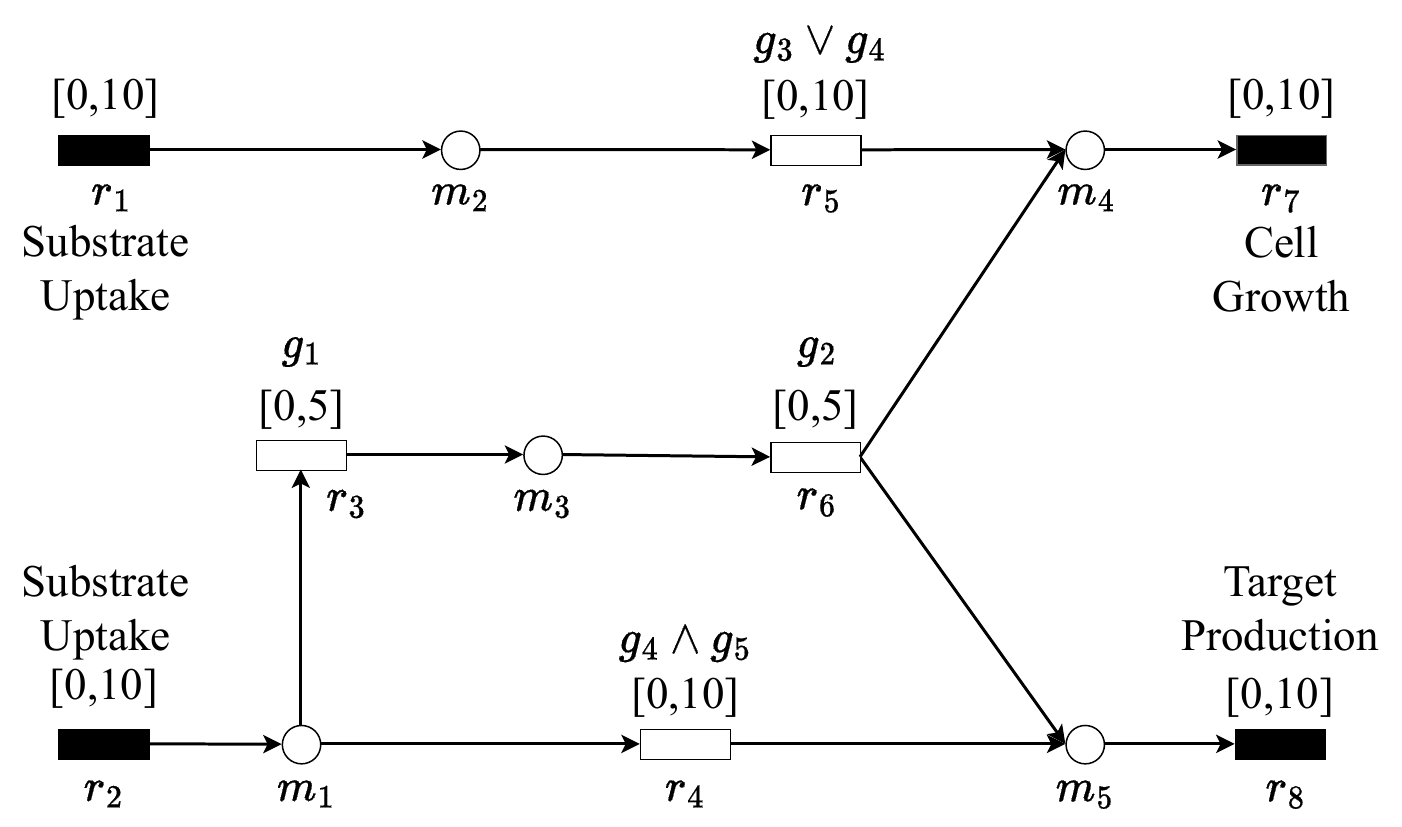}
  \caption{A toy example of the constraint-based model where circles and rectangles represent metabolites and reactions, respectively. 
  Black and white rectangles denote external and internal reactions, respectively. 
  $r_1$, $r_2$ correspond to two substrate uptake reactions.
  $r_7$, $r_8$ correspond to cell growth, and target metabolite production reactions, respectively. 
  The reaction rates are constrained by the range $[l_i, u_i]$.
  In this example, all stoichiometric ratios are 1 or -1.}  
  \label{fig:toy_model_1}
\end{figure}

\subsubsection{\textbf{Toy constraint-based model setup}}
\label{subsubsec:toy model}
Fig. \ref{fig:toy_model_1} shows a toy example in which $\mathcal{C} = \{\textit{M}, \textit{R}, \textit{S}, \textit{L}, \textit{U}, \textit{G}, \textit{F}, \textit{P}\}$ represents a constraint-based model.
Specifically, for its metabolic network part $\mathcal{C}_1$, we have $\textit{M} = \{m_1, \ldots, m_5\}$, $\textit{R} = \{r_1, \ldots, r_8\}$, $\textit{S} = \begin{pmatrix}
  1  & 0  & -1 & -1 & 0  & 0  & 0  & 0   \\
  0  & 1  & 0  & 0  & -1 & 0  & 0  & 0    \\
  0  & 0  & 1  & 0  & 0  & -1 & 0  & 0    \\
  0  & 0  & 0  & 0  & 1  & 1  & -1 & 0    \\
  0  & 0  & 0  & 1  & 0  & 1  & 0  & -1   \\
\end{pmatrix}$, 
$\textit{L} = \{0, 0, 0, 0, 0, 0, 0, 0\}$, and $\textit{U} = \{10, 10, 5, 10, 10, 5, 10, 10\}$.
For its gene reaction rule part $\mathcal{C}_2$, we have $G=\{g_1,\ldots,g_5\}$, $F=\{f_1,\ldots,f_8\}$ and $P=\{p_1,\ldots,p_8\}$, where $f_i:p_i=1$ for $i\in\{1,2,7,8\}$; $f_3:p_3=g_1$; $f_4:p_4=g_4\wedge g_5$; $f_5:p_5=g_3\vee g_4$; and $f_6:p_6=g_2$.

In this toy model $\mathcal{C}$, the gene--reaction rule for $r_4$ is given as \(f_4: \; p_4 = g_4 \wedge g_5\).
Thus, the reaction rate of $r_4$ (denoted $v_4$) is forced to be $0$ if either $g_4$ or $g_5$ is $0$ (deleted). 
If both $g_4$ and $g_5$ are $1$ (not deleted), then \(0 \leq v_4 \leq 10 \).
For $r_5$, the reaction rate $v_5$ is forced to be $0$ only if both $g_3$ and $g_4$ are $0$. 
If at least one of $g_3$ or $g_4$ is $1$, then \(0 \leq v_5 \leq 10 \).
For $r_1$, $r_2$, $r_7$, and $r_8$, since $p_1 = p_2 = p_7 = p_8 = 1$ always holds, none of $v_1$, $v_2$, $v_7$, or $v_8$ can be forced to $0$ by any gene deletions.

In this example, $m_5$ is designated as the target metabolite, produced through the target metabolite production reaction $r_8$.

\begin{table}
  \centering
  \caption{Gene deletion strategies for the toy example in Fig. \ref{fig:toy_model_1} are classified into five types based on the resulting flux (reaction rate) distributions.}
  \label{table: toy_1_sum}
  \resizebox{0.99\linewidth}{!}{%
  \begin{tabular}{cl} 
  \toprule
  \textbf{Type} & \textbf{Gene deletion strategies classified by flux distributions} \\ 
  \midrule
  1    & $\emptyset$, \{$g_1$\}, \{$g_2$\}, \{$g_3$\}, \{$g_1$, $g_2$\}, \{$g_1$, $g_3$\}, \{$g_2$, $g_3$\}, \{$g_1$, $g_2$, $g_3$\}\\
  2    & \{$g_4$\}, \{$g_5$\}, \{$g_3$, $g_5$\}, \{$g_4$, $g_5$\}\\
  3    & \{$g_1$, $g_4$\}, \{$g_2$, $g_4$\}, \{$g_1$, $g_2$, $g_4$\}, \{$g_1$, $g_5$\}, \{$g_2$, $g_4$\}, \{$g_1$, $g_2$, $g_5$\}, \\
       & \{$g_1$, $g_3$, $g_5$\}, \{$g_1$, $g_4$, $g_5$\}, \{$g_2$, $g_3$, $g_5$\}, \{$g_2$, $g_4$, $g_5$\}, \\
       & \{$g_1$, $g_2$, $g_3$, $g_5$\},  \{$g_1$, $g_2$, $g_4$, $g_5$\}\\
  4    & \{$g_3$, $g_4$\}, \{$g_3$, $g_4$, $g_5$\} \\
  5    & \{$g_1$, $g_3$, $g_4$\}, \{$g_2$, $g_3$, $g_4$\}, \{$g_1$, $g_2$, $g_3$, $g_4$\}, \{$g_1$, $g_3$, $g_4$, $g_5$\}\\
       & \{$g_2$, $g_3$, $g_4$, $g_5$\}, \{$g_1$, $g_2$, $g_3$, $g_4$, $g_5$\}\\
  \bottomrule
  \end{tabular}
  }
\end{table}

\begin{table}
\centering
\caption{The resulting flux distributions for gene deletion strategy Types 1 to 5. For each type, the most optimistic (best) and pessimistic (worst) target metabolite production rate (PR, $v_8$) at cell growth rate (GR, $v_7$) maximization are provided.}
\label{table: toy_1_fd}
 \resizebox{0.99\linewidth}{!}{%
\begin{tabular}{ccccccccccc} 
\toprule
\multirow{2}{*}{\textbf{ID}} & \multirow{2}{*}{\begin{tabular}[c]{@{}c@{}}\textbf{Deletion}\\\textbf{strategy}\end{tabular}} & \multirow{2}{*}{\begin{tabular}[c]{@{}c@{}}\textbf{PR}\\\textbf{situation}\end{tabular}} & \multicolumn{8}{c}{\textbf{Flux distribution}}                                                                                              \\ 
\cmidrule{4-11}
                             &                                                                                               &                                                                                          & \textbf{$v_1$} & \textbf{$v_2$} & \textbf{$v_3$} & \textbf{$v_4$} & \textbf{$v_5$} & \textbf{$v_6$} & \textbf{$v_7$}      & \textbf{$v_8$}  \\ 
\midrule
1                            & \multirow{2}{*}{Type 1}                                                                       & best                                                                                     & 10             & 10             & 0              & 10             & 10             & 0              & \multirow{2}{*}{10} & 10              \\
2                            &                                                                                               & worst                                                                                    & 10             & 0              & 0              & 0              & 10             & 0              &                     & 0               \\ 
\midrule
3                            & \multirow{2}{*}{Type 2}                                                                       & best                                                                                     & 5              & 5              & 5              & 0              & 5              & 5              & \multirow{2}{*}{10} & 5               \\
4                            &                                                                                               & worst                                                                                    & 10             & 0              & 0              & 0              & 10             & 0              &                     & 0               \\ 
\midrule
5                            & \multirow{2}{*}{Type 3}                                                                       & best                                                                                     & 10             & 0              & 0              & 0              & 10             & 0              & \multirow{2}{*}{10} & 0               \\
6                            &                                                                                               & worst                                                                                    & 10             & 0              & 0              & 0              & 10             & 0              &                     & 0               \\ 
\midrule
7                            & \multirow{2}{*}{Type 4}                                                                       & best                                                                                     & 0              & 5              & 5              & 0              & 0              & 5              & \multirow{2}{*}{5}  & 5               \\
8                            &                                                                                               & worst                                                                                    & 0              & 5              & 5              & 0              & 0              & 5              &                     & 5               \\ 
\midrule
9                            & \multirow{2}{*}{Type 5}                                                                       & best                                                                                     & 0              & 0              & 0              & 0              & 0              & 0              & \multirow{2}{*}{0}  & 0               \\
10                           &                                                                                               & worst                                                                                    & 0              & 0              & 0              & 0              & 0              & 0              &                     & 0               \\
\bottomrule
\end{tabular}
}
\end{table}

\subsubsection{\textbf{Gene deletions on the toy constraint-based model}}
\label{subsubsec:toy model deletion}
Table \ref{table: toy_1_sum} describes the patterns of gene deletions in this example: the $2^5 = 32$ patterns are classified into five types based on the flux distributions. 
The detailed flux distributions for all types can be found in Table \ref{table: toy_1_fd}.

In the original toy model without gene deletions, two paths lead to cell growth (\(r_7\)): \(r_1 \rightarrow r_5 \rightarrow r_7\) and \(r_2 \rightarrow r_3 \rightarrow r_6 \rightarrow r_7\). 
Two paths from \(r_2\) reach the target production reaction (\(r_8\)): \(r_2 \rightarrow r_3 \rightarrow r_6 \rightarrow r_8\) and \(r_2 \rightarrow r_4 \rightarrow r_8\). 
A maximal growth rate of \(GR = 10\) is achievable, but the flux distribution is not unique.

If both the path leading to cell growth \((r_1 \rightarrow r_5 \rightarrow r_7)\) and the path leading to target production \((r_2 \rightarrow r_4 \rightarrow r_8)\) are used at maximum flux, then \(GR = 10\) and \(PR = 10\) are obtained, as shown in the flux distribution of ID~1 in Table~\ref{table: toy_1_fd}. 
This represents the most optimistic case for the flux value of PR. 
In contrast, if only the path \((r_1 \rightarrow r_5 \rightarrow r_7)\) is used at maximum flux, then \(GR = 10\) and \(PR = 0\) are obtained, as shown in the flux distribution of ID 2 in Table~\ref{table: toy_1_fd}. 
This represents the most pessimistic case for the flux value of PR. 
As previously defined, in the context of growth-coupled production, we evaluate PR in the most pessimistic case when GR is maximized. 
Therefore, in the original state of the toy model, we have \(GR = 10\) and \(PR = 0\). 
We denote this flux distribution as Type 1. 
Seven gene deletion strategies, \(\{g_1\}\), \(\{g_2\}\), \(\{g_3\}\), \(\{g_1, g_2\}\), \(\{g_1, g_3\}\), \(\{g_2, g_3\}\), and \(\{g_1, g_2, g_3\}\), are also classified as Type 1, as they yield the same flux distribution.

IDs 7 and 8 in Table~\ref{table: toy_1_fd} describe the flux distributions when \(g_3\) and \(g_4\) are deleted. 
In this case, \(f_4: p_4 = g_4 \wedge g_5 = 0\) and \(f_5: p_5 = g_3 \vee g_4 = 0\), forcing \(v_4 = 0\) and \(v_5 = 0\). 
This yields a uniquely determined flux distribution with \(v_2 = v_3 = v_6 = 4\) and \(v_1 = v_4 = v_5 = 0\). 
Consequently, both the optimistic and pessimistic PR values are 5, with maximal \(GR = 5\). 
We denote this flux distribution as Type 4. 
Another gene deletion strategy, \(\{g_3, g_4, g_5\}\), is also classified as Type~4, as shown in Table~\ref{table: toy_1_fd}.

The detailed analysis of the flux distribution under deletion strategies Type 2, 3, and 5 can be found in the Appendix.

Across the five types of gene deletion strategies analyzed above, only Type~4 satisfies \(PR \geq PR_{\text{threshold}}\) and \(GR \geq GR_{\text{threshold}}\), even in the most pessimistic scenario. 
Consequently, in this toy model, only the two Type~4 gene deletion strategies, \(\{g_3, g_4\}\) and \(\{g_3, g_4, g_5\}\), enable growth-coupled production of the target metabolite \(m_5\).

\subsection{Problem Setting}
\label{subsec:Problem}
\textbf{Gene deletion strategy prediction.} Given a constraint-based metabolic model \( \mathcal{C} \) and a set of metabolites \( \textit{M} \), the task of gene deletion strategy prediction seeks to infer the gene deletion strategies \( \textit{DS} \) of target metabolites \( \textit{M}_{\text{target}} \subseteq \textit{M} \). 
This task operates under the assumption that gene deletion strategies for a subset of metabolites, denoted as \( \textit{DS}_{\text{known}} \), are already established, while the strategies for the target metabolites remain unknown.
The primary objective is to develop a computational agent, such as a predictive model \( \mathcal{F}(\theta) \), which processes the known deletion strategies along with the information on the metabolic model.
This agent aims to generate predictions \( \widetilde{\textit{DS}} \) for the target metabolites, optimizing its outputs to closely approximate the ground-truth deletion strategies \( \textit{DS}^{*} \), as presented below:

\begin{algorithm}
  \caption*{Prediction of Gene Deletion Strategies for Target Metabolites.}
  \textbf{Given:} \( \mathcal{C} \), \( {\textit{DS}}_{\text{known}} \) \\
  \textbf{Objective:} A computational agent \( \mathcal{F}(\theta) \), parameterized by \( \theta \), to generate deletion strategy predictions \( \widetilde{\textit{DS}} \) for the target metabolites \( \textit{M}_{\text{target}} \subseteq \textit{M} \). \par\vspace{-0.5em}
  \begin{flushleft}
    \hspace{1em}\textbf{Such that:} \\
    \hspace{2em}\( \widetilde{{\textit{DS}}} = \mathcal{F}(\theta; {\textit{DS}}_{\text{known}}, \mathcal{C}) \); \\
    \hspace{2em}\( \widetilde{\textit{DS}} \approx {\textit{DS}}^{*} \)
  \end{flushleft}
  \label{algo: Gene Deletion Prediction}
\end{algorithm}

\section{Method}
\label{sec:method}
In this section, we first introduce our method for constructing graph representations from input metabolic models.
Next, we present the \textit{\proposed} framework, outlining its overall design, explaining the structure and function of each module, and describing the specific deep learning models used for predicting gene deletion strategies.
Finally, we provide implementation details, including the parameter settings and training procedures of the \textit{\proposed} framework.

\subsection{Constructing Graph Representations of Metabolic Models}
\label{subsec:graph}
We present a pipeline for constructing a biologically meaningful graph representation from a genome-scale metabolic model. 
The process begins by extracting core biochemical entities and their interactions, followed by two refinement steps that incorporate both graph-structural properties and domain-specific metabolite knowledge. 
The resulting graph preserves essential metabolic structure while reducing confounding topological noise, thereby enabling more interpretable and effective downstream graph-based learning.

The construction procedure consists of the following four steps:

\begin{itemize}
  \item \textbf{Step 1: Construct Temporary Graph form Input Metabolic Model.} \\
  We begin with a genome-scale metabolic model as input.
  As the metabolic model lacks an explicit topological form, we draft a temporary graph \(\mathcal{G}_{temp}\) from its components.   
  We construct \(\mathcal{G}_{temp}\) as a tripartite heterogeneous graph consisting of three node types: metabolites, reactions, and genes. 
  Directed edges are defined from metabolites to reactions (for substrates), from reactions to metabolites (for products), and from genes to reactions (based on GPR rules). 
  This graph encodes the complete structural and regulatory logic of the metabolic model.

  \item \textbf{Step 2: Induce the Metabolite Graph.} \\
  To focus on metabolite-level connectivity, we remove gene nodes and treat each reaction node in \(\mathcal{G}_{temp}\) as a hyperedge connecting all its associated metabolites. 
  We then project this hypergraph into a pairwise undirected graph by linking each reactant to each product within the same reaction. 
  The resulting metabolite graph, denoted as \(\mathcal{G}_0\), contains only metabolite nodes, with edges representing biochemical transformations.

  \item \textbf{Step 3: Graph Attribute-Based Node Filtering.} \\
  Highly connected hub-like metabolites can distort the network’s topological structure by disproportionately influencing properties such as shortest paths and information flow.
  To reduce this topological noise, we identify and remove from \(\mathcal{G}_0\) the ten metabolites with the highest degree, resulting in a trimmed graph \(\mathcal{G}_1\).

  \item \textbf{Step 4: Knowledge-Based Node Refinement.} \\
  We apply a two-step refinement process to enhance biological interpretability, resulting in the final graph \(\mathcal{G}_{\text{final}}\):

  \textbf{(1) Remove canonical currency metabolites.} 
  We exclude known currency metabolites based on a curated list from KEGG and BiGG databases, including: 
  \texttt{["atp", "adp", "amp", "nad", "nadh", "nadp", "nadph", "coa", "h", "h2o", "pi", "ppi", "co2", "nh4", "o2", "glutathione", "na1", "k", "mg2", "ca2"]}.
  \textbf{(2) Selectively restore removed metabolites.} 
  Any metabolite removed in previous steps that corresponds to a designated production target (e.g., a growth-coupled product) or a non-currency, high-degree metabolite is selectively reintroduced to maintain the functional integrity of the \(\mathcal{G}_{\text{final}}\).
\end{itemize}

As a result, the proposed graph construction pipeline produces a representation that preserves key functional structures while minimizing the confounding influence of topological hubs.
The pseudocode of the proposed graph construction pipeline is presented below.

\begin{algorithm}
  \caption*{Constructing Metabolite Graph from a Constraint-Based Metabolic Model}

  \textbf{Input:} 
  \(
  \mathcal{C} = \{\textit{M}, \textit{R}, \textit{S}, \textit{G}, \textit{F} \}
  \),  Currency Metabolite Set $\mathcal{U}$.

  \begin{flushleft}
  \textbf{Step 1: Construct Temporary Graph from Input Metabolic Model} \\
  Define node set:
  \[
    V_{\text{temp}} = \textit{M} \cup \textit{R} \cup \textit{G}
  \]
  Define directed edge set \(E_{\text{temp}}\):
  \begin{itemize}
    \item Add \((m_i, r_j)\) if \(S_{ij} < 0\) (metabolite is a substrate)
    \item Add \((r_j, m_i)\) if \(S_{ij} > 0\) (metabolite is a product)
    \item Add \((g_k, r_j)\) if \(g_k\) appears in \(f_j\) (GPR rule)
  \end{itemize}
  Resulting graph: 
  \[
    \mathcal{G}_{\text{temp}} = (V_{\text{temp}}, E_{\text{temp}})
  \]

  \textbf{Step 2: Induce Metabolite Graph \(\mathcal{G}_0\)} \\
  % Remove gene nodes; only metabolites remain as nodes.
  Project each reaction node into undirected metabolite pairs:
  \begin{itemize}
    \item For each reaction \(r_j\), for every substrate \(m_i\) (\(S_{ij} < 0\)) and every product \(m_{i'}\) (\(S_{i'j} > 0\)), add undirected edge \((m_i, m_{i'})\).
  \end{itemize}
  Define metabolite-only graph:
  \[
    \mathcal{G}_0 = (V_0 = \textit{M},\ E_0)
  \]

  \textbf{Step 3: Graph Attribute-Based Node Filtering} \\
  Compute degrees:
  \[
    \deg(m) = |\{v \mid (m, v) \in E_0,\ m, v \in V_0 \}|
  \]
  Select top-\(k\) high-degree nodes:
  \[
    H_k = \text{Top-}k \text{ metabolites by degree}
  \]
  Remove them:
  \[
    V_1 = V_0 \setminus H_k,\quad E_1 = \{ (m_i, m_j) \in E_0 \mid m_i, m_j \in V_1 \}
  \]
  \[
    \mathcal{G}_1 = (V_1,\ E_1)
  \]

  \textbf{Step 4: Knowledge-Based Node Refinement} \\
  \textbf{(1) Remove canonical currency metabolites given by $\mathcal{U}$}:
  \[
    V_2 = V_1 \setminus \mathcal{U},\quad E_2 = \{ (m_i, m_j) \in E_1 \mid m_i, m_j \in V_2 \}
  \]
  \[
    \mathcal{G}_2 = (V_2,\ E_2)
  \]

  \textbf{(2) Selectively restore removed metabolites}:
  \[
    T = \{ m \in H_k \mid m = m_{\text{target}} \text{ or } m \notin \mathcal{U} \}
  \]
  \[
    V_3 = V_2 \cup T
  \]
  \[
    E_T = \{ (m_i, m_j) \in E_0 \mid (m_i \in T \text{ or } m_j \in T) \text{ and } m_i, m_j \in V_3 \}
  \]
  \[
    E_3 = E_2 \cup E_T
  \]
  \[
    \mathcal{G}_{\text{final}} = (V_3,\ E_3)
  \]

  \textbf{Output:} Final graph \(\mathcal{G}_{\text{final}} = (V_3, E_3)\)
  \end{flushleft}
  \label{algo:graph_construction}
\end{algorithm}

\subsection{\textit{\proposed} Framework Overview}

\begin{figure*}[t]
  \centering
  \includegraphics[width=0.9\linewidth]{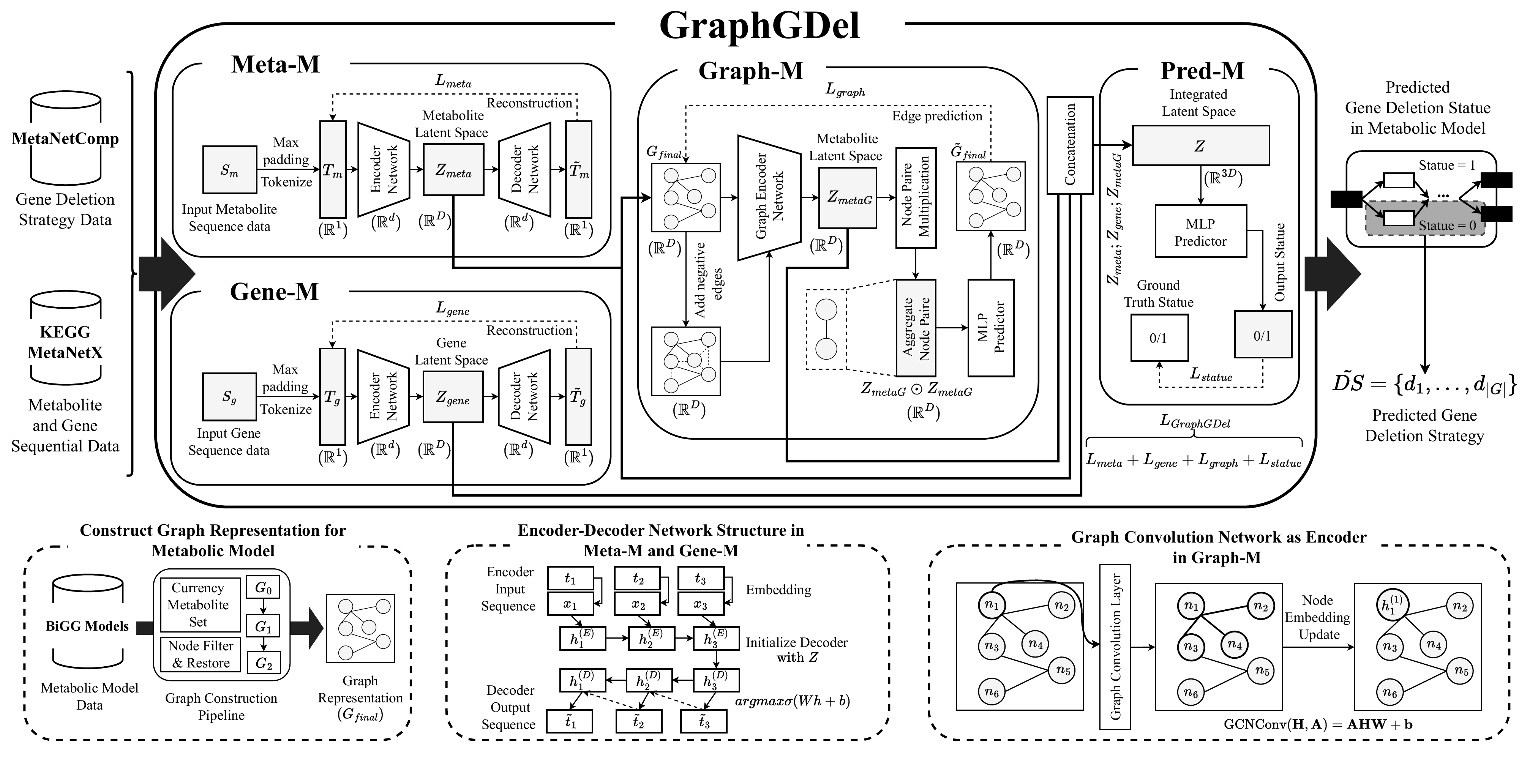}
  \caption{A system overview of the proposed gene deletion strategy prediction framework. The framework comprises four neural network-based modules: (1) \textbf{Meta-M}, which learns the metabolite latent representation $Z_{meta}$, (2) \textbf{Gene-M}, which learns the gene latent representation $Z_{gene}$, (3) \textbf{Graph-M}, which learns the refined metabolite latent representation $Z_{metaG}$ in a specific metabolic graph, and (4) \textbf{Pred-M}, which integrates the three upstream latents into a new latent representation, $Z$, for final gene deletion prediction.}  
  \label{fig:framework}
\end{figure*}

Metabolic processes are determined not only by the intrinsic properties of genes and metabolites but also by the complex interactions among metabolites. 
While sequence information captures molecular-level characteristics of genes and metabolites, it is the upstream and downstream relationships between metabolites that determine how perturbations, such as gene deletions, propagate through the system. 
Capturing both perspectives is therefore desirable for accurately predicting growth-coupled production outcomes.

To address this, the \textit{\proposed} framework integrates sequence learning with graph-based learning to achieve:  
(1) extraction of gene and metabolite features from sequence data using sequence learning;  
(2) refinement of metabolite features in the context of the metabolic network using graph-based learning; and  
(3) integration of sequence- and graph-based features to model gene–metabolite interdependencies for prediction.  

Specifically, as illustrated in Fig.~\ref{fig:framework}, \textit{\proposed} consists of four neural network modules: the metabolite representation learning module (\textbf{Meta-M}), the gene representation learning module (\textbf{Gene-M}), the graph representation learning module (\textbf{Graph-M}), and the deletion strategy prediction module (\textbf{Pred-M}).

\begin{itemize}
\item \textbf{Meta-M}:  
This module captures the characteristics of metabolites in the metabolic model.  
It formulates a metabolite representation learning task that projects input metabolite data into a latent representation, ${Z}_{\text{meta}}$, summarizing metabolite-relevant information.  
This unsupervised learning task typically employs a sequence learning model with an autoencoder structure.  
The learned ${Z}_{\text{meta}}$ serves as a foundation for both downstream graph-based learning and the final gene deletion prediction task.  

\item \textbf{Gene-M}:  
This module captures the characteristics of genes in the metabolic model.  
Parallel to Meta-M, it formulates a gene representation learning task that projects input gene data into a latent representation, ${Z}_{\text{gene}}$, summarizing gene-relevant information.  
This unsupervised learning task can also employ a sequence learning model with an autoencoder structure.  
The learned ${Z}_{\text{gene}}$ serves as a foundation for the final gene deletion prediction task.  

\item \textbf{Graph-M}:  
This module captures the topological relationships and structural patterns within the metabolic graph to refine metabolite features.  
It formulates a link prediction task that maps the input metabolic graph and metabolite features into a refined latent representation, ${Z}_{\text{metaG}}$, integrating both metabolite characteristics and their graph connectivity patterns.  
This unsupervised learning task employs a graph learning model as the encoder and a pairwise node aggregation mechanism as the decoder.  
The learned ${Z}_{\text{metaG}}$ provides metabolic model-specific information for the final gene deletion prediction task.
 
\item \textbf{Pred-M}:  
This module integrates the latent representations ${Z}_{\text{meta}}$ from Meta-M, ${Z}_{\text{gene}}$ from Gene-M, and ${Z}_{\text{metaG}}$ from Graph-M into a unified latent representation ${Z}$, capturing interdependencies between metabolites and genes.  
The integration process combines sequence-based and graph-based features, transforming them into discriminative features optimized for prediction.  
Finally, a supervised predictor is trained on ${Z}$ to infer the deletion status of each gene for different target metabolites within the given metabolic model.  
\end{itemize}

\subsection{Gene Deletion Strategy Prediction}

\subsubsection{\textbf{Metabolite Representation Learning}}

The Meta-M module is designed to transform raw character strings input, i.e., the SMILES (Simplified Molecular Input Line Entry System) sequences of metabolites into dense, informative feature vectors \cite{smiles}. 
To effectively capture the complex patterns within these chemical representations, we employ a sequence-to-sequence autoencoder architecture based on Long Short-Term Memory (LSTM) networks \cite{lstm_original,seq2seq}.
We selected LSTM as the default model because our goal is to propose a general learning framework for this task and to validate its effectiveness. 
The LSTM meets these requirements: it is straightforward to implement and provides a strong baseline for future comparisons with more complex replacements, such as Transformer-based models \cite{Transformer}.

Given a SMILES sequence $\mathbf{S}_m$ of length $L_m$, the module first performs character-level tokenization to convert the chemical string into a numerical representation suitable for neural network processing.
The tokenization process constructs a vocabulary $\mathcal{V}$ from all unique characters appearing in the SMILES dataset, including atomic symbols, bond indicators, and structural notations. 
Each SMILES sequence $\mathbf{S}_m^{(i)}$ is then mapped to a sequence of token indices:
\begin{equation}
    \mathbf{T}_m^{(i)} = [t_1, t_2, \dots, t_{L_m}], \quad t_j \in \mathcal{V},
\end{equation}
where $t_j$ represents the index of the $j$-th character in the vocabulary.

To ensure consistent dimensionality across different sequence lengths, we apply zero-padding to standardize all sequences to a maximum length $L_{\max}$:
\begin{equation}
    \mathbf{T}_m^{(i)} = [t_1, t_2, \dots, t_{L_m}, 0, 0, \dots, 0], \quad \text{if } L_m < L_{\max}.
\end{equation}

The tokenized sequence is then embedded into a continuous vector space through a learnable embedding layer:
\begin{equation}
    \mathbf{X}_m = \text{Embedding}(\mathbf{T}_m),
\end{equation}
where the embedding operation maps each token index to a dense vector representation of dimension $d$.

The embedded sequence is processed through an LSTM encoder that captures the sequential dependencies and chemical patterns within the SMILES string:
\begin{equation}
    \mathbf{h}_l = \text{LSTM}(\mathbf{x}_l, \mathbf{h}_{l-1}, \mathbf{c}_{l-1}),
\end{equation}
where $\mathbf{h}_l \in \mathbb{R}^d$ represents the hidden state at time step $l$, and $\mathbf{c}_l$ is the cell state that maintains long-term memory.

To stabilize the training process and improve generalization, we apply layer normalization to the LSTM outputs:
\begin{equation}
    \mathbf{h}_l^{\text{norm}} = \text{LayerNorm}(\mathbf{h}_l).
\end{equation}

The final metabolite representation $\mathbf{Z}_{\text{meta}} \in \mathbb{R}^D$ is obtained through temporal pooling of the normalized hidden states:
\begin{equation}
    \mathbf{Z}_{\text{meta}} = \frac{1}{L_m} \sum_{l=1}^{L_m} \mathbf{h}_l^{\text{norm}}.
\end{equation}

To enable unsupervised learning, the module includes a decoder component that reconstructs the original SMILES sequence from the learned representation. 
The decoder employs an LSTM network that takes the encoded representation $\mathbf{Z}_{\text{meta}}$ as input and generates a sequence of token predictions:
\begin{equation}
    \tilde{\mathbf{T}}_m = \text{Decoder}(\mathbf{Z}_{\text{meta}}).
\end{equation}

The training objective for Meta-M combines reconstruction loss with representation learning:
\begin{equation}
    \mathcal{L}_{\text{meta}} = \frac{1}{L_m} \sum_{l=1}^{L_m} \text{CrossEntropy}(\tilde{\mathbf{t}}_l, \mathbf{t}_l),
\end{equation}
where $\mathbf{t}_l$ and $\tilde{\mathbf{t}}_l$ are the ground-truth and predicted tokens at position $l$, respectively.

Through this autoencoder architecture, Meta-M learns to capture the fundamental chemical characteristics of metabolites such as functional groups and structural patterns, resulting in a compact representation $\mathbf{Z}_{\text{meta}} \in \mathbb{R}^D$ that preserves the molecular information necessary for downstream graph-based learning and prediction tasks.

\subsubsection{\textbf{Gene Representation Learning}}

The Gene-M module employs an identical neural network architecture to Meta-M, but processes gene sequences represented as amino acid sequences instead of SMILES strings. 

Given a gene sequence input $\mathbf{S}_g$ of length $L_g$, the module follows the same processing pipeline as Meta-M:
\begin{equation}
    \mathbf{T}_g^{(i)} = [t_1, t_2, \dots, t_{L_g}], \quad t_j \in \mathcal{V}_g,
\end{equation}
where $\mathcal{V}_g$ contains the 20 standard amino acid characters and any additional symbols.

The identical LSTM autoencoder architecture processes the amino acid sequence through the following network layers:
\begin{equation}
  \left\{
  \begin{aligned}
    \mathbf{X}_g &= \text{Embedding}(\mathbf{T}_g), \\
    \mathbf{h}_l &= \text{LSTM}(\mathbf{x}_l, \mathbf{h}_{l-1}, \mathbf{c}_{l-1}), \\
    \mathbf{h}_l^{\text{norm}} &= \text{LayerNorm}(\mathbf{h}_l), \\
    \mathbf{Z}_{\text{gene}} \in \mathbb{R}^D &= \frac{1}{L_g} \sum_{l=1}^{L_g} \mathbf{h}_l^{\text{norm}}, \\
    \tilde{\mathbf{T}}_g &= \text{Decoder}(\mathbf{Z}_{\text{gene}}),
  \end{aligned}
  \right.
\end{equation}

The training objective follows the same reconstruction loss formulation:
\begin{equation}
    \mathcal{L}_{\text{gene}} = \frac{1}{L_g} \sum_{l=1}^{L_g} \text{CrossEntropy}(\tilde{\mathbf{t}}_l, \mathbf{t}_l),
\end{equation}
where the LSTM is trained to extract informative patterns from amino acid sequences.

This parallel design ensures balanced feature extraction from both biological entities, providing a unified foundation for the subsequent integration in the prediction module.

\subsubsection{\textbf{Graph Representation Learning}}

To capture the topological relationships and structural patterns within the metabolic network, we design Graph-M, which employs Graph Convolutional Networks (GCN) to refine the metabolite representations learned by Meta-M in the context of the metabolic network structure. 
This module takes the sequence-based metabolite features from Meta-M as initial node embeddings and further enhances them by incorporating the graph connectivity patterns through graph convolutions.

Given a metabolic graph \( \mathcal{G} = (V, E) \) where \( V \) represents the set of metabolite nodes and \( E \) represents the set of edges connecting metabolites through reactions, Graph-M processes the graph structure using the Meta-M learned features as initial node representations. 
Specifically, for each metabolite node \( v_i \in V \), we initialize its node features using the corresponding Meta-M representation \( \mathbf{Z}_{\text{meta}}^{(i)} \in \mathbb{R}^D \).

The initial node feature matrix \( \mathbf{X} \in \mathbb{R}^{|V| \times D} \) is constructed by aggregating the Meta-M representations for all metabolites:
\begin{equation}
    \mathbf{X} = [\mathbf{Z}_{\text{meta}}^{(1)}, \mathbf{Z}_{\text{meta}}^{(2)}, \ldots, \mathbf{Z}_{\text{meta}}^{(|V|)}]^T,
\end{equation}
where each row \( \mathbf{X}_i \) corresponds to the Meta-M learned representation of metabolite \( v_i \).

To enhance the node features with structural information, we add learnable positional embeddings that capture node identity within the graph:
\begin{equation}
    \mathbf{X}_{\text{enhanced}} = \mathbf{X} + \mathbf{P},
\end{equation}
where \( \mathbf{P} \in \mathbb{R}^{|V| \times d_{\text{pos}}} \) represents learnable positional embeddings for each node, and \( d_{\text{pos}} \) is the positional embedding dimension.

The enhanced node features are then processed through two graph convolutional layers to capture both local and global structural patterns:
\begin{align}
    \mathbf{H}^{(1)} &= \text{ReLU}(\text{GCNConv}(\mathbf{X}_{\text{enhanced}}, \mathbf{A})), \\
    \mathbf{H}^{(2)} &= \text{GCNConv}(\mathbf{H}^{(1)}, \mathbf{A}),
\end{align}
where \( \mathbf{A} \) is the adjacency matrix of the graph, and \( \text{GCNConv}(\cdot, \cdot) \) represents the graph convolutional operation defined as:
\begin{equation}
    \text{GCNConv}(\mathbf{H}, \mathbf{A}) = \mathbf{A} \mathbf{H} \mathbf{W} + \mathbf{b},
\end{equation}
where \( \mathbf{W} \) and \( \mathbf{b} \) are learnable weight matrix and bias vector, respectively.

The final graph-refined metabolite embeddings \( \mathbf{Z}_{\text{metaG}} \in \mathbb{R}^{|V| \times D} \) are obtained from the second graph convolutional layer:
\begin{equation}
    \mathbf{Z}_{\text{metaG}} = \mathbf{H}^{(2)}.
\end{equation}

To train the Graph-M module, we employ a link prediction task that predicts the existence of edges between node pairs. 
For a given node pair \( (i, j) \), we compute the link prediction score as:
\begin{equation}
    s_{ij} = \sigma(\mathbf{W}_{\text{link}}[\mathbf{Z}_{\text{metaG}}^{(i)} \| \mathbf{Z}_{\text{metaG}}^{(j)}] + \mathbf{b}_{\text{link}}),
\end{equation}
where \( \| \) denotes concatenation, \( \mathbf{W}_{\text{link}} \) and \( \mathbf{b}_{\text{link}} \) are learnable parameters, and \( \sigma(\cdot) \) is the sigmoid activation function.

The training objective for Graph-M is the binary cross-entropy loss for link prediction:
\begin{equation}
    \mathcal{L}_{\text{graph}} = -\sum_{(i,j) \in E} \log(s_{ij}) - \sum_{(i,j) \notin E} \log(1 - s_{ij}),
\end{equation}
where the first term corresponds to positive edges (existing connections) and the second term corresponds to negative edges (non-existing connections).

By minimizing this loss, Graph-M learns to refine the Meta-M learned metabolite representations by incorporating the topological relationships within the metabolic network. 
The resulting \( \mathbf{Z}_{\text{metaG}} \) provides graph-enhanced metabolite features that combine the intrinsic molecular characteristics captured by Meta-M with the structural context of the metabolic network, thereby offering a more comprehensive representation for the downstream prediction task.

\subsubsection{\textbf{Gene Deletion Status Prediction}}

The Pred-M module integrates the learned representations from all three upstream modules to predict gene deletion status. 
Given the metabolite representation \( \mathbf{Z}_{\text{meta}} \) from Meta-M, the gene representation \( \mathbf{Z}_{\text{gene}} \) from Gene-M, and the graph-refined metabolite representation \( \mathbf{Z}_{\text{metaG}} \) from Graph-M, the integration process combines these features to create a unified representation for prediction.

For a specific gene-metabolite pair, we select the corresponding metabolite node from the graph embeddings:
\begin{equation}
    \mathbf{Z}_{\text{metaG}}^{(m)} = \mathbf{Z}_{\text{metaG}}[m],
\end{equation}
where \( m \) is the index of the target metabolite in the graph.

The three representations are then concatenated to form a combined feature vector:
\begin{equation}
    \mathbf{Z} = [\mathbf{Z}_{\text{meta}} \| \mathbf{Z}_{\text{gene}} \| \mathbf{Z}_{\text{metaG}}^{(m)}],
\end{equation}
where \( \| \) denotes concatenation, resulting in a feature vector of dimension \( 3D \).

The combined representation is then processed through a fully connected layer followed by a sigmoid activation to produce the final prediction:
\begin{equation}
    \hat{y} = \sigma(\mathbf{W}_{\text{pred}} \mathbf{Z} + \mathbf{b}_{\text{pred}}),
\end{equation}
where \( \mathbf{W}_{\text{pred}} \in \mathbb{R}^{1 \times 3D} \) and \( \mathbf{b}_{\text{pred}} \in \mathbb{R} \) are learnable parameters, and \( \sigma(\cdot) \) is the sigmoid activation function.

The output \( \hat{y} \in [0, 1] \) represents the probability that the gene should be deleted (when \( \hat{y} \) is close to 0) or retained (when \( \hat{y} \) is close to 1) for the given metabolite.

The training objective for the prediction task is the binary cross-entropy loss:
\begin{equation}
    \mathcal{L}_{\text{pred}} = -y \log(\hat{y}) - (1 - y) \log(1 - \hat{y}),
    \label{eq:pred_loss}
\end{equation}
where \( y \in \{0, 1\} \) is the ground-truth label indicating whether the gene should be deleted (0) or retained (1).

The overall training objective for the entire \textit{\proposed} framework combines all four loss components:
\begin{equation}
    \mathcal{L}_{\text{total}} = \mathcal{L}_{\text{meta}} + \mathcal{L}_{\text{gene}} + \mathcal{L}_{\text{graph}} + \mathcal{L}_{\text{pred}}.
\end{equation}

This multi-task learning approach ensures that each module learns meaningful representations while contributing to the final prediction task. 
The framework is trained end-to-end, allowing the representations from each module to be optimized jointly for the gene deletion prediction task.

\subsection{Implementation Details and Model Training}
\subsubsection{Device}
All neural network models were trained within the proposed framework on an NVIDIA A100-PCIE-40GB GPU with a Xeon Gold 6258R CPU. 
Model parameters were optimized for 100 epochs using the Adam optimizer with a learning rate of $1\times 10^{-3}$. 

\subsubsection{Data Partitioning Strategy}
Model performance was evaluated using repeated stratified 10-fold cross-validation. The dataset was randomly partitioned into 10 folds while preserving class proportions. In each fold, approximately 90\% of the data were used for training and 10\% for validation.

This procedure was repeated five times with different random splits, resulting in a total of $5 \times 10$ cross-validation runs. All reported performance metrics correspond to these repeated cross-validation experiments and are presented as the mean $\pm$ standard deviation across folds.

\subsubsection{Hyperparameters}\label{subsubsec:hyperparams}
Hyperparameters were tuned via grid search within the repeated 10-fold stratified cross-validation framework (5 repeats with different random seeds) on the training/validation data. 
The configuration achieving the highest average validation performance across folds was selected as the final setting. 
The detailed search space and results are summarized in the Appendix.

\section{Computational Experiments}
\label{sec:experiments}
In this section, we first provide an overview of our design for computational experiments on metabolic models. 
Next, we describe the experimental setup in detail, including the data, baseline methods, and evaluation metrics employed. 
Finally, we present the experimental results along with key observations.

\subsection{Overview}
\label{subsec:ExpOverview} 

In the computational experiments, we consider three metabolic models: \textit{e\_coli\_core}, \textit{iMM904}, and \textit{iML1515}:

\begin{itemize}
  \item
  \textbf{\textit{e\_coli\_core}} \cite{orth2010reconstruction} is a simplified metabolic model of \textit{Escherichia coli} (E. coli) that captures only the essential pathways necessary for core metabolic functions. 
  It is widely used for foundational studies of bacterial metabolism and gene interactions.
  \item
  \textbf{\textit{iMM904}} \cite{mo2009connecting} is a more comprehensive model of \textit{Saccharomyces cerevisiae} (yeast), incorporating a larger set of reactions and metabolites. 
  It is commonly used for analyzing complex eukaryotic metabolic networks.
  \item
  \textbf{\textit{iML1515}} \cite{monk2017ml1515} is one of the most detailed metabolic models of E. coli, covering a vast array of metabolic reactions and pathways. 
  It provides a more intricate network compared to \textit{e\_coli\_core} and is widely used as a genome-scale model for simulations.
\end{itemize}

We further define several key terms employed in our computational experiments as follows:

\begin{itemize}
  \item
  \textbf{Growth-essential genes}: These are genes whose deletion results in a maximum GR of zero during FBA. They are crucial for cell growth and should not be deleted when aiming for growth-coupled production.
  
  \item
  \textbf{Pre-determined genes}: A gene is defined as pre-determined if its deletion state is consistent (either deleted or non-deleted) for more than 95\% of metabolites within the metabolic model. Such genes' deletion states are considered stable and will not be included in the evaluation.

  \item
  \textbf{GCP-possible metabolites}: These metabolites are theoretically capable of achieving growth-coupled production within the metabolic models. Only GCP-possible metabolites are considered suitable for targeting when calculating gene deletion strategies.
\end{itemize}

The detailed breakdown of genes, metabolites, and reactions in the studied models is summarized in Table \ref{table: model_sum}.

\begin{table}
  \centering
  \caption{Summary of constraint-based metabolic models used in the computational experiments.}
  \label{table: model_sum}
  \resizebox{0.99\linewidth}{!}{%
  \begin{tabular}{lccc} 
  \toprule
  \textbf{Metabolic model}                & \textbf{\textit{e\_coli\_core}}           & \textbf{\textit{iMM904}}                    & \textbf{\textit{iML1515}}  \\ 
  \midrule
  \#Genes              & 137                     & 905                       & 1516     \\ 
  \#Growth essential genes                       & 7             & 110    & 196      \\
  \#Pre-determined genes & 71         & 850    & 1483      \\
  \midrule
  \#Metabolites        & 72                      & 1226                      & 1877     \\
  \#GCP-possible metabolites & 48                      & 782                       & 1085     \\ 
  \midrule
  \#Reactions          & 95                      & 1577                      & 2712     \\
  \bottomrule
  \end{tabular}
  }
\end{table}

To evaluate the effectiveness and robustness of the proposed framework for graph representation construction and gene deletion strategy prediction, we design three computational experiments on the aforementioned metabolic models:

\begin{enumerate}
  \item 
  \textbf{Performance Evaluation:} We assess gene deletion strategy prediction methods, including the proposed method and baseline approaches, on gene deletion strategy prediction tasks. 
  All methods are applied to the three metabolic models, and results are reported separately for each model.

  \item 
  \textbf{Modules Ablation:} We investigate the contribution of different learning modules by selectively removing specific components from the proposed framework. 
  Two ablation variants, Ge-SMILES (using only Gene-M and Meta-M) and Ge-Graph (using only Gene-M and Graph-M), are evaluated across the three metabolic models, with results reported separately for each model.

  \item 
  \textbf{Graph Representation Evaluation:} We assess the impact of different graph representations on the proposed framework for gene deletion strategy prediction. 
  Two graph representations, $\mathcal{G}_0$ and $\mathcal{G}_1$ (introduced in Section~\ref{subsec:graph}), generated during the early stages of the graph construction pipeline, are used as input to the learning model. 
  Experiments are conducted on the three metabolic models, with results reported separately for each model.
\end{enumerate}

\subsection{Data and Preprocessing}
\label{subsec:data}
For the experiments, we use data from four categories associated with each metabolic model: (1) gene deletion strategy data, which serves as the ground truth for prediction tasks; (2) SMILES data, providing information on the metabolites in the models; (3) amino acid (AA) sequence data, providing information on the genes, specifically their encoded proteins in the models; and (4) the metabolite network structure. 

\subsubsection{\textbf{Gene Deletion Strategy Data}}
\label{subsec:deletion_strategy_data}

\textbf{Description.} 
In this study, gene deletion strategy data are represented as binary sequences that encode deletion strategies for specific metabolites.
Specifically, the gene deletion strategy data \( \textit{DS} = (d_1, \ldots, d_{\lvert \textit{G} \rvert}) \) is a binary sequence of length \( \lvert \textit{G} \rvert \), where \( \lvert \textit{G} \rvert \) is the number of genes in the metabolic model, and each element \( d_i \in \{0,1\} \) indicates the state of gene \( i \): \( d_i = 0 \) if gene \( i \) is deleted, and \( d_i = 1 \) if it is retained.
A single target metabolite may have multiple deletion strategies, as more than one strategy can achieve growth-coupled production. 
In this study, we focus on the maximal gene deletion strategy, denoted \( \textit{DS}_{\text{max}} \), where deleting any additional gene would abolish the growth-coupled production of the metabolite.

\textbf{Data Source.} 
The gene deletion strategy data used in this study are obtained from the MetNetComp database \cite{tamura2023metnetcomp}.
MetNetComp curates systematically simulated results derived from constraint-based models and computes growth-coupled gene deletion strategies that are minimal or maximal with respect to the number of deleted genes.

We download all available maximal gene deletion strategy data for the three metabolic models under investigation: \textit{e\_coli\_core}, \textit{iMM904}, and \textit{iML1515}. 
The raw data include four types of information: (1) the number of remaining genes, (2) the maximum growth rate, (3) the minimum production rate, and (4) a list of remaining genes. 
We use only the fourth category, which specifies the genes retained in the maximal gene deletion strategy. 
Binary gene deletion sequences can be readily derived from this remaining gene list.

\subsubsection{\textbf{SMILES Data}}
\label{subsec:SMILES data}

\textbf{Data Source.}
The SMILES data used in this study is obtained from the MetaNetX database \cite{MetaNetX}. 
MetaNetX is an online platform for accessing, analyzing, and manipulating genome-scale metabolic networks and biochemical pathways. 
It curates information from public repositories to integrate data on metabolites and related reactions across many species.
For all three metabolic models under study, we download SMILES data for metabolites corresponding to those with available gene deletion strategy data. 
For metabolites lacking corresponding SMILES entries in the database, we use dummy SMILES sequences as placeholders.

\subsubsection{\textbf{Amino Acid Sequence Data}}
\label{subsec:AA data}

\textbf{Data Source.}
The AA sequence data used in this study is obtained from the KEGG database \cite{KEGG}. 
KEGG provides comprehensive information on protein-coding genes, their products, and associated metabolic pathways. 
The AA sequence data in KEGG is curated from publicly available genomic resources and is continuously updated to ensure accuracy and relevance.
We download AA sequence data for all three metabolic models under study, covering all genes in each model. 
For genes lacking corresponding AA sequences in the database, we use dummy sequences as placeholders.

To mitigate class imbalance between deleted and non-deleted genes, we first applied gene-wise balancing: for genes with both deletion states, an equal number of samples was randomly drawn from each state; otherwise, all available instances were retained. 
Next, global balancing was performed by downsampling the majority class to match the minority, resulting in a fully balanced training set. 
Due to metabolite number constraints, the final sample sizes approximate, rather than exactly match, the intended ratios.
The details of the constructed datasets for the three metabolic models are summarized in Table \mbox{\ref{table: data_sum}}.

\begin{table}
  \centering
  \caption{Summary of constructed datasets for three constraint-based metabolic models.}
  \label{table: data_sum}
  \resizebox{0.99\linewidth}{!}{%
  \begin{tabular}{lccc} 
\toprule
\textbf{Metabolic model}     & \textbf{\textit{e\_coli\_core}} & \textbf{\textit{iMM904}} & \textbf{\textit{iML1515}}  \\ 
\midrule
\#Deletion strategies        & 41                              & 105                      & 443                        \\
\#Genes per strategy         & 137                             & 905                      & 1516                       \\ 
\midrule
\#SMILES terms               & 41                              & 99                       & 425                        \\
\#Dummy SMILES terms         & 0                               & 6                        & 18                         \\ 
\midrule
\#AA Seq. terms              & 136                             & 889                      & 1513                       \\
\#Dummy AA Seq. terms        & 1                               & 16                       & 3                          \\ 
\midrule
\#Balanced train/val samples & 4932                            & 85070                    & 603368                     \\
\#Balanced test samples      & 685                             & 9955                     & 68220                      \\
\bottomrule
\end{tabular}
}
\end{table}

\subsection{Baseline Method: DNN}
\label{subsec:baseline_dnn}
In the first set of experiments, i.e., performance evaluation of gene deletion strategy prediction methods, we employed a deep learning–based baseline: a Deep Neural Network (DNN) \cite{DNN} to predict gene deletion status.
DNNs are widely used as standard baselines in supervised learning tasks because they can capture non-linear relationships in high-dimensional data, even with relatively simple architectures.

In our case, the DNN baseline takes as input the same tokenized and padded
sequences of the gene ${\mathbf{T}_g}$ and the metabolite ${\mathbf{T}_m}$
as used in the proposed framework.
Unlike the modular, task-specific architecture of our method, this baseline
adopts a simpler design: ${\mathbf{T}_g}$ and ${\mathbf{T}_m}$ are
independently mapped to dense vectors through separate embedding layers
and then concatenated to form a joint representation following:
\[
\begin{aligned}
\mathbf{X}_g &= \mathrm{Embedding}(\mathbf{T}_g), \\
\mathbf{X}_m &= \mathrm{Embedding}(\mathbf{T}_m), \\
\mathbf{X}   &= [\mathbf{X}_g ; \mathbf{X}_m].
\end{aligned}
\]
The concatenated embedding ${\mathbf{X}}$ is passed through two fully connected
layers with ReLU activations, followed by a sigmoid output layer that produces
the prediction score.
% \begin{equation}
% \left\{
% \begin{aligned}
% \mathbf{h}_1 &= \mathrm{ReLU}(\mathbf{W}_1 \mathbf{X} + \mathbf{b}_1), \\
% \hat{y}      &= \sigma(\mathbf{W}_2 \mathbf{h}_1 + \mathbf{b}_2),
% \end{aligned}
% \right.
% \end{equation}
% where $\hat{y} = P(y = 1 \mid \mathbf{T}_g, \mathbf{T}_m; \theta)$,
% $y \in \{0, 1\}$ is the ground-truth label (deleted or non-deleted),
% and $\theta = \{\mathbf{W}_1,\mathbf{W}_2,\mathbf{b}_1,\mathbf{b}_2\}$ denotes the set of learnable parameters.
The DNN is trained with the same experimental setup as our proposed framework,
using the binary cross-entropy loss defined in Equation~\ref{eq:pred_loss}.

This DNN baseline provides a point of comparison for a naive learnable classifier that does not incorporate any task-specific design. 
Its performance helps evaluate the necessity of our proposed modular design, which integrates distinct tasks (i.e., representation learning in Gene-M, Meta-M, and Grpah-M) within the framework.

\subsection{Baseline Method: DeepGdel}
\label{subsec:baseline_DeepGdel}
Besides the DNN baseline, we also employ DeepGdel~\cite{yang2025deepgdel} as an advanced baseline for gene deletion strategy prediction.
DeepGdel extends simple DNN approaches by incorporating sequential learning through LSTM-based autoencoders for both gene and metabolite representations, with details introduced in the Appendix.

DeepGdel is trained using a multi-task learning objective that combines the reconstruction losses from the autoencoders with the binary classification loss.
This baseline provides a more sophisticated comparison, demonstrating the effectiveness of sequential learning approaches and highlighting the additional benefits of incorporating graph structure information in our proposed framework.

\subsection{Baseline Method: NMA (Neighborhood Mean Aggregation)}
\label{subsec:baseline_NMA}
To evaluate the benefit of learned graph representations over a simple topology-aware scheme, we include an NMA (Neighborhood Mean Aggregation) baseline.
NMA uses the same Gene-M and Meta-M modules as DeepGdel and the same metabolite graph structure as in the proposed framework (Section~\ref{subsec:graph}).
For prediction, instead of using the metabolite's own Meta-M embedding, NMA aggregates Meta-M outputs over the graph neighborhood: for metabolite node $i$ with neighbor set $\mathcal{N}(i)$,
\[
\mathbf{z}_i^{\mathrm{nma}} = \frac{1}{|\mathcal{N}(i)|} \sum_{j \in \mathcal{N}(i)} \mathbf{z}_j^{\mathrm{meta}},
\]
where $\mathbf{z}_j^{\mathrm{meta}}$ is the Meta-M embedding of metabolite $j$; if $|\mathcal{N}(i)| = 0$, $\mathbf{z}_i^{\mathrm{nma}} = \mathbf{z}_i^{\mathrm{meta}}$.
The (gene, metabolite) prediction is then
\[
\hat{y} = \sigma\bigl(\mathbf{w}^\top (\mathbf{h}_g \odot \mathbf{z}_i^{\mathrm{nma}}) + b\bigr),
\]
with $\mathbf{h}_g$ the Gene-M embedding and $\odot$ the element-wise product.
NMA is trained using the same protocol as the other baselines and serves as a topology-aware, non-learned graph aggregation baseline for comparison with our graph-learning method.

\subsection{Evaluation Metrics}
\label{subsec:metrics}
To ensure a fair comparison, the same data partitioning strategy and hyperparameter tuning protocol (grid search with 10-fold stratified cross-validation, repeated five times with different random seeds) were applied to the proposed framework and to all baseline methods (DNN, DeepGdel, and NMA) in the performance evaluation experiments; the best configuration for each method was selected based on validation performance.
We evaluate model performance using four metrics: \textbf{Overall Accuracy}, \textbf{Macro-averaged Precision, Recall, F1 Score}, and \textbf{AUC}. 
The detailed definitions and formulations of these metrics can be found in the Appendix.

\begin{table*}[t]
  \centering
  \caption{Comparison of gene deletion strategy prediction performance between the proposed GraphGDel framework and baseline methods (DNN, DeepGdel, NMA) across three metabolic models. “M-Avg.” denotes macro-averaged results. Best-performing values are shown in bold.}
  \label{table: main_exp}
  \resizebox{0.98\linewidth}{!}{%
\begin{tabular}{lcccccccccccc} 
\toprule
\multirow{2}{*}{Metric}       & \multicolumn{4}{c}{e\_coli\_core}                   & \multicolumn{4}{c}{iMM904}                          & \multicolumn{4}{c}{iML1515}                          \\ 
\cmidrule(l){2-5} \cmidrule(l){6-9} \cmidrule(l){10-13}
                              & DNN               & DeepGdel & NMA    & Proposed                  & DNN               & DeepGdel & NMA    & Proposed                  & DNN               & DeepGdel & NMA    & Proposed                  \\ 
\midrule
Overall Accuracy (\%) & 63.31 $\pm$ 2.70 & 71.18 $\pm$ 0.44 & 72.25 $\pm$ 0.52 & \textbf{77.35~$\pm$ 1.45} & 69.28 $\pm$ 1.58 & 80.58 $\pm$ 0.44 & 81.18 $\pm$ 0.50 & \textbf{85.54 $\pm$ 1.04} & 60.04 $\pm$ 1.04 & 67.91 $\pm$ 0.34 & 68.52 $\pm$ 0.38 & \textbf{73.22 $\pm$ 1.02} \\
M-Avg. Precision (\%) & 64.99 $\pm$ 2.57 & 72.44 $\pm$ 0.56 & 73.12 $\pm$ 0.58 & \textbf{75.99 $\pm$ 1.33} & 72.07 $\pm$ 1.44 & 77.51 $\pm$ 0.42 & 78.38 $\pm$ 0.45 & \textbf{83.19 $\pm$ 1.07} & 64.60 $\pm$ 1.13 & 68.90 $\pm$ 0.35 & 69.48 $\pm$ 0.38 & \textbf{74.67 $\pm$ 1.09} \\
M-Avg. Recall (\%)    & 64.13 $\pm$ 2.20 & 72.87 $\pm$ 0.54 & 73.42 $\pm$ 0.56 & \textbf{75.38 $\pm$ 1.37} & 68.52 $\pm$ 1.36 & 77.85 $\pm$ 0.44 & 78.52 $\pm$ 0.48 & \textbf{82.91 $\pm$ 1.12} & 61.80 $\pm$ 1.11 & 64.40 $\pm$ 0.35 & 65.18 $\pm$ 0.38 & \textbf{72.51 $\pm$ 1.06} \\
M-Avg. F1 Score (\%)  & 59.59 $\pm$ 2.36 & 70.85 $\pm$ 0.65 & 71.52 $\pm$ 0.68 & \textbf{74.45 $\pm$ 1.24} & 64.84 $\pm$ 1.56 & 76.48 $\pm$ 0.45 & 77.28 $\pm$ 0.48 & \textbf{81.87 $\pm$ 1.25} & 58.69 $\pm$ 1.17 & 63.22 $\pm$ 0.34 & 64.08 $\pm$ 0.36 & \textbf{72.31 $\pm$ 1.03} \\
M-Avg. AUC (\%)       & 62.66 $\pm$ 2.89 & 71.16 $\pm$ 0.85 & 72.08 $\pm$ 0.88 & \textbf{76.56 $\pm$ 1.26} & 66.50 $\pm$ 1.47 & 80.24 $\pm$ 0.78 & 81.02 $\pm$ 0.82 & \textbf{84.24 $\pm$ 1.12} & 60.30 $\pm$ 1.14 & 65.78 $\pm$ 0.54 & 66.52 $\pm$ 0.56 & \textbf{73.17 $\pm$ 1.04} \\    
\bottomrule
\end{tabular}
}
\end{table*}

To provide additional insight into model behavior and error distribution, we report confusion matrices for the proposed method and the baseline methods (DNN, DeepGdel, and NMA) on the three metabolic models in Table~\ref{table:confusion}.
Table~\ref{table:confusion} reports row-normalized proportions for each true label. The positive class corresponds to ``non-deleted'' (gene not deleted) and the negative class to ``deleted.''

\begin{table}[t]
  \centering
  \caption{Row-normalized confusion matrices of all methods across three metabolic models. Rows correspond to true labels and columns to predicted labels. Values are row-wise proportions.}
  \label{table:confusion}
  \resizebox{\linewidth}{!}{%
  \begin{tabular}{llcccccc}
  \toprule
  \multirow{3}{*}{Method}
  & \multirow{3}{*}{\shortstack{True \\ label}}
  & \multicolumn{2}{c}{\textit{e\_coli\_core}}
  & \multicolumn{2}{c}{\textit{iMM904}}
  & \multicolumn{2}{c}{\textit{iML1515}} \\
  \cmidrule(lr){3-4} \cmidrule(lr){5-6} \cmidrule(lr){7-8}
  &
  & \shortstack{Pred. \\ Del.}
  & \shortstack{Pred. \\ N-Del.}
  & \shortstack{Pred. \\ Del.}
  & \shortstack{Pred. \\ N-Del.}
  & \shortstack{Pred. \\ Del.}
  & \shortstack{Pred. \\ N-Del.} \\
  \midrule

  \multirow{2}{*}{DNN}
  & Del.   & 0.69 & 0.31 & 0.71 & 0.29 & 0.55 & 0.45 \\
  & N-Del. & 0.42 & 0.58 & 0.34 & 0.66 & 0.31 & 0.69 \\
  \midrule

  \multirow{2}{*}{DeepGdel}
  & Del.   & 0.76 & 0.24 & 0.88 & 0.12 & 0.78 & 0.22 \\
  & N-Del. & 0.34 & 0.66 & 0.33 & 0.67 & 0.49 & 0.51 \\
  \midrule

  \multirow{2}{*}{NMA}
  & Del.   & 0.79 & 0.21 & 0.90 & 0.10 & 0.77 & 0.23 \\
  & N-Del. & 0.33 & 0.67 & 0.30 & 0.70 & 0.40 & 0.60 \\
  \midrule

  \multirow{2}{*}{Proposed}
  & Del.   & 0.83 & 0.17 & 0.93 & 0.07 & 0.75 & 0.25 \\
  & N-Del. & 0.32 & 0.68 & 0.27 & 0.73 & 0.30 & 0.70 \\
  \bottomrule
  \end{tabular}%
  }
\end{table}

\begin{table*}[t]
\centering
\caption{Ablation study of the proposed framework comparing two variants: Ge-SMILES (using only gene and SMILES-based metabolite features) and Ge-Graph (using only gene and graph-based metabolite features). Results are evaluated across the three metabolic models under study. “M-Avg.” denotes macro-averaged results. Best-performing values are shown in bold.}
\label{table: ablation_exp_1}
\resizebox{0.98\linewidth}{!}{%
\begin{tabular}{lcccccccccc} 
\toprule
\multirow{2}{*}{Metric}       & \multicolumn{3}{c}{e\_coli\_core}            & \multicolumn{3}{c}{iMM904}                   & \multicolumn{3}{c}{iML1515}                   \\ 
\cmidrule{2-10}
                              & Ge-SMILES    & Ge-Graph  & Proposed          & Ge-SMILES    & Ge-Graph  & Proposed          & Ge-SMILES    & Ge-Graph  & Proposed          \\ 
\midrule
Overall Accuracy (\%)         & 70.34 $\pm$ 1.25 & 74.33 $\pm$ 1.35 & \textbf{77.35 $\pm$ 1.45} & 78.76 $\pm$ 1.17 & 83.33 $\pm$ 1.38 & \textbf{85.54 $\pm$ 1.04} & 67.11 $\pm$ 0.96 & 70.35 $\pm$ 1.38 & \textbf{73.22 $\pm$ 1.02} \\
M-Avg. Precision (\%)         & 70.34 $\pm$ 1.26 & 72.72 $\pm$ 1.55 & \textbf{75.99 $\pm$ 1.33} & 78.34 $\pm$ 1.12 & 81.74 $\pm$ 1.37 & \textbf{83.19 $\pm$ 1.07} & 68.45 $\pm$ 0.94 & 71.49 $\pm$ 1.36 & \textbf{74.67 $\pm$ 1.09} \\
M-Avg. Recall (\%)            & 71.95 $\pm$ 1.24 & 72.31 $\pm$ 1.36 & \textbf{75.38 $\pm$ 1.37} & 77.29 $\pm$ 1.12 & 81.36 $\pm$ 1.49 & \textbf{82.91 $\pm$ 1.12} & 63.23 $\pm$ 0.88 & 69.30 $\pm$ 1.35 & \textbf{72.19 $\pm$ 0.91} \\
M-Avg. F1 Score (\%)          & 70.23 $\pm$ 1.22 & 71.24 $\pm$ 1.54 & \textbf{74.45 $\pm$ 1.24} & 77.44 $\pm$ 1.15 & 80.93 $\pm$ 1.22 & \textbf{81.87 $\pm$ 1.25} & 62.13 $\pm$ 0.93 & 69.45 $\pm$ 1.45 & \textbf{72.31 $\pm$ 1.03} \\
M-Avg. AUC (\%)               & 70.37 $\pm$ 1.32 & 73.47 $\pm$ 1.37 & \textbf{76.56 $\pm$ 1.26} & 78.49 $\pm$ 1.15 & 82.42 $\pm$ 1.41 & \textbf{84.24 $\pm$ 1.12} & 68.35 $\pm$ 0.93 & 70.44 $\pm$ 1.39 & \textbf{73.17 $\pm$ 1.04} \\ 
\bottomrule
\end{tabular}
}
\end{table*}

\begin{table*}[t]
  \centering
  \caption{Performance comparison of the proposed framework using different graph representations ($\mathcal{G}_0$ and $\mathcal{G}_1$) across three metabolic models. “M-Avg.” denotes macro-averaged results. The best-performing values are shown in bold.}
  \label{table: ablation_exp_2}
  \resizebox{0.98\linewidth}{!}{%
\begin{tabular}{lccccccccc} 
\toprule
\multirow{2}{*}{Metric}       & \multicolumn{3}{c}{e\_coli\_core}                   & \multicolumn{3}{c}{iMM904}                          & \multicolumn{3}{c}{iML1515}                          \\ 
\cmidrule(l){2-4} \cmidrule(l){5-7} \cmidrule(l){8-10}
                              & $\mathcal{G}_0$               & $\mathcal{G}_1$               & Proposed                  & $\mathcal{G}_0$               & $\mathcal{G}_1$               & Proposed                  & $\mathcal{G}_0$               & $\mathcal{G}_1$               & Proposed                  \\ 
\midrule
Overall Accuracy (\%) & 73.25 $\pm$ 1.45 & 76.20 $\pm$ 1.50 & \textbf{77.35~$\pm$ 1.45} & 82.02 $\pm$ 1.04 & 84.70 $\pm$ 1.05 & \textbf{85.54 $\pm$ 1.04} & 70.15 $\pm$ 1.02 & 72.80 $\pm$ 1.00 & \textbf{73.22 $\pm$ 1.02} \\
M-Avg. Precision (\%) & 73.10 $\pm$ 1.33 & 75.85 $\pm$ 1.30 & \textbf{75.99 $\pm$ 1.33} & 80.33 $\pm$ 1.07 & 83.05 $\pm$ 1.05 & \textbf{83.19 $\pm$ 1.07} & 70.88 $\pm$ 1.09 & 73.92 $\pm$ 1.10 & \textbf{74.67 $\pm$ 1.09} \\
M-Avg. Recall (\%)    & 73.55 $\pm$ 1.37 & \textbf{75.45 $\pm$ 1.38} & 75.38 $\pm$ 1.37 & 81.01 $\pm$ 1.12 & \textbf{83.10 $\pm$ 1.10} & 82.91 $\pm$ 1.12 & 68.42 $\pm$ 0.91 & \textbf{72.51 $\pm$ 0.93} & 72.19 $\pm$ 0.91 \\
M-Avg. F1 Score (\%)  & 72.67 $\pm$ 1.24 & 74.40 $\pm$ 1.20 & \textbf{74.45 $\pm$ 1.24} & 80.57 $\pm$ 1.25 & 81.75 $\pm$ 1.20 & \textbf{81.87 $\pm$ 1.25} & 68.09 $\pm$ 1.03 & 72.20 $\pm$ 1.05 & \textbf{72.31 $\pm$ 1.03} \\
M-Avg. AUC (\%)       & 73.12 $\pm$ 1.26 & 76.10 $\pm$ 1.20 & \textbf{76.56 $\pm$ 1.26} & 81.55 $\pm$ 1.12 & 83.70 $\pm$ 1.10 & \textbf{84.24 $\pm$ 1.12} & 69.67 $\pm$ 1.04 & 72.85 $\pm$ 1.02 & \textbf{73.17 $\pm$ 1.04} \\    
\bottomrule
\end{tabular}
}
\end{table*}

\subsection{Results and Performance Comparison}
\label{subsec:results}
The results of the three computational experiments, namely (1) Performance Evaluation, (2) Module Ablation, and (3) Graph Representation Evaluation, are summarized in Tables~\ref{table: main_exp}, \ref{table: ablation_exp_1}, and \ref{table: ablation_exp_2}, respectively.
To assess whether the reported differences are statistically significant, we performed paired $t$-tests on the cross-validation fold-level results (50 paired values per comparison, obtained from 5 repeats $\times$ 10 folds), comparing the proposed method against each baseline. All comparisons yielded $p < 0.05$ across every metric and model (Tables~\ref{table: main_exp}, \ref{table: ablation_exp_1}, and \ref{table: ablation_exp_2}).

\textbf{Performance Evaluation:}
Across all metabolic models, \textit{\proposed} consistently outperformed the baseline methods in overall accuracy, macro-averaged precision, recall, F1 score, and AUC.
Specifically, the overall accuracy of \textit{\proposed} was 77.35\%, 85.54\%, and 73.22\% for \textit{e\_coli\_core}, \textit{iMM904}, and \textit{iML1515}, respectively, representing substantial improvements over DNN (63.31\%, 69.28\%, 60.04\%), DeepGdel (71.18\%, 80.58\%, 67.91\%), and NMA (72.25\%, 81.18\%, 68.52\%).
This demonstrates the robustness of \textit{\proposed} across models of varying scales.

\textbf{Modules Ablation:}
To assess the contributions of the learning modules, we evaluated two variants of \textit{\proposed}: Ge-SMILES (using only Gene-M and Meta-M) and Ge-Graph (using only Gene-M and Graph-M).
Across all models, relying solely on SMILES-based or graph-based metabolite features in combination with gene features consistently resulted in reduced performance.
For example, in \textit{e\_coli\_core}, the overall accuracy dropped from 77.35\% (full model) to 70.34\% (Ge-SMILES) and 74.33\% (Ge-Graph).
These results highlight that integrating both sequential and graph-based metabolite representations is critical for optimal performance.

\textbf{Graph Representation Evaluation:}
We further examined the impact of different graph representations generated at early stages of the pipeline ($\mathcal{G}_0$ and $\mathcal{G}_1$).
Across all metabolic models, using the early-stage representations resulted in decreased performance compared to the proposed graph representation.
For instance, in \textit{iMM904}, the overall accuracy was 82.02\% for $\mathcal{G}_0$, 84.70\% for $\mathcal{G}_1$, and 85.54\% for the proposed representation.
This demonstrates the importance of the proposed graph construction pipeline for capturing meaningful metabolic relationships.

Overall, the experimental results confirm that \textit{\proposed} effectively integrates sequential and graph-based representations, leveraging the constructed graph to enhance gene deletion strategy prediction. 
The improvements are consistent across models of different sizes, indicating the generalizability and robustness of the framework.

\section{Discussion and Conclusion}
\label{sec:discussion}
In this study, we address the gene deletion strategy prediction problem by introducing a graph-based representation approach.
Specifically, we (1) propose a systematic pipeline for constructing graph representations from genome-scale metabolic models, and (2) develop \textit{\proposed}, a learning framework that integrates sequential representations with graph-structural information to enhance gene deletion strategy prediction in genome-scale metabolic models.

In the computational experiments, we compared \textit{\proposed} with three baseline methods (DNN, DeepGdel, and NMA) on three constraint-based metabolic models of different scales.  
The results demonstrate the effectiveness of \textit{\proposed} for predicting gene deletion strategies across models of varying sizes, consistently outperforming all baseline methods across all five evaluation metrics.  
Specifically, the overall accuracy of \textit{\proposed} for \textit{e\_coli\_core}, \textit{iMM904}, and \textit{iML1515} was 77.35\%, 85.54\%, and 73.22\%, respectively, representing improvements over the DNN baseline (14.04\%, 16.26\%, and 13.18\% higher), the DeepGdel baseline (6.17\%, 4.96\%, and 5.31\% higher), and the NMA baseline (5.10\%, 4.36\%, and 4.70\% higher).

We also conducted ablation studies to assess the impact of different learning modules on gene deletion strategy prediction performance.  
Using only SMILES-based metabolite features (Ge-SMILES) or graph-based metabolite features (Ge-Graph) resulted in consistent performance degradation across all metabolic models.  
These results highlight that integrating both sequential and graph-based metabolite representations is crucial for achieving optimal performance in gene deletion strategy prediction.
Furthermore, our graph representation evaluation experiments highlight the importance of the proposed graph representations in graph-based learning tasks for metabolic models.  
The proposed framework, when learning from these graph representations, consistently achieves the best performance, whereas using representations from the early stages of the construction pipeline ($\mathcal{G}_0$ and $\mathcal{G}_1$) results in performance degradation.  
These findings demonstrate the value of a carefully designed graph representation construction pipeline.

The proposed graph representation construction pipeline generates graphs from genome-scale metabolic models by leveraging both graph-structural properties and metabolite knowledge to filter biochemical interactions, thereby enabling downstream graph-based learning tasks.  
Since genome-scale metabolic models encode more complex relationships among elements, such as GPR rules linking genes to reactions, the successful application of the proposed metabolite graph suggests the potential to integrate multiple types of relationships to further enhance gene deletion strategy prediction.  
For instance, the pipeline could be extended to construct multi-layered graphs where different node types represent metabolites, genes, and reactions, with edges capturing various relationships, including biochemical transformations and GPR rules.

A key innovation of \textit{\proposed} is the explicit incorporation of graph structure information through the Graph-M module, addressing a limitation of previous approaches that relied solely on sequential data.  
By leveraging metabolic network topology, \textit{\proposed} captures relationships between metabolites that are not apparent from their sequential representations.  
While this work uses LSTM autoencoders and Graph Convolutional Networks as the default architecture to demonstrate and benchmark the framework's feasibility, its modular design allows for the integration of more advanced models.  
For example, one potential improvement is to incorporate recent large language models specialized for molecular or protein sequences into the learning modules of \textit{\proposed}.
A limitation of this study is that the ground-truth labels are derived from constraint-based simulations curated in the MetNetComp database. Validation against independent biological experiments or alternative re-simulation frameworks was beyond the scope of the current work, and extending the evaluation to such settings remains an important direction for future research.
Furthermore, the current framework does not explicitly provide pathway-level or graph-level explanations of its predictions. Incorporating interpretability techniques, such as attention mechanisms or graph attribution methods to highlight influential subgraphs or pathways, represents an important direction for future work.
In this study, we focused on maximal gene deletion strategies. In the reference database, minimal deletion strategies are derived from maximal deletion sets; therefore, maximal deletions contain richer structural information and provide a more comprehensive representation of the feasible perturbation space. 
Importantly, the proposed framework is not inherently restricted to maximal deletions. Because it operates on general gene perturbation patterns, it can be readily extended to minimal or experimentally constrained intervention strategies. Exploring such practically constrained settings will be an important direction for future work.

In future work, we plan to enhance the graph representation construction pipeline by incorporating additional types of model elements, such as reactions and genes. 
We also aim to benchmark more advanced deep learning architectures for metabolite and gene representation learning across metabolic models from a broader range of species.

\section{ACKNOWLEDGMENTS}
This work was supported in part by JSPS, KAKENHI under Grant 23K20386, 26K02972, and JST SPRING, Grant Number JPMJSP2110.

\bibliographystyle{naturemag}
\bibliography{Bibliography}

\end{document}